\documentclass[fleqn,twoside]{article}
\usepackage[headings]{espcrc2}

\readRCS
$Id: espcrc2.tex,v 1.2 2004/02/24 11:22:11 spepping Exp $
\usepackage{graphicx, balance}
\usepackage{amsmath}
\usepackage{amssymb}
\usepackage{amsfonts}
\usepackage{tikz}
\usepackage{textcomp}
\usepackage{mathtools}
\usepackage{setspace}
\usepackage{fancyhdr}
\usepackage{mathptmx,bm}
\usepackage{enumerate}
\usepackage[figuresright]{rotating}
\newtheorem{thrm}{Theorem}
\newtheorem{defn}{Definition}
\newtheorem{prop}{Proposition}

\title{\textbf{Threshold Multi Secret Sharing Using Elliptic Curve and Pairing}}

\author{V P Binu\address[DCA]{Department of Computer Applications, Cochin University  of Science and Technology, \\~Cochin-682022 India, Contact: binuvp@gmail.com \\},
A  Sreekumar\address{Associate Professor, Department of Computer Applications, Cochin University  of Science and Technology, Cochin-22, India}}

\runtitle{Threshold Multi Secret Sharing Using Elliptic Curve and Pairing}
\runauthor{Binu V P, et al.,}

\begin{document}
\begin{abstract}

Secret Sharing techniques are now the building blocks of several security protocols. A $(t,n)$ threshold secret sharing scheme is one in which $t$ or more participant can join together to retrieve the secret.Traditional single secret sharing schemes are modified and generalized to share multiple secrets.Use of elliptic curve and pairing in secret sharing is gaining more importance.In this paper we propose a threshold multi secret sharing scheme where more than one secret is shared.When the threshold number of participants collate, the multi secret can be retrieved. The scheme make use of elliptic curve and bilinear pairing.Verification of shares by the participants, shares consistency checking, detection and identification of cheaters are the extended capabilities achieved.The shared secrets are  retrieved in single stage here, unlike the multi stage secret sharing scheme.The participants can be added very easily.The scheme is efficient and the number of public parameters are also less compared with the existing threshold multi secret sharing scheme based on the elliptic curve.The dealer can easily modify the secret or add additional secret by changing the public parameters of the scheme.This is the first proposal of a threshold  multi secret sharing scheme with extended capabilities using self pairing.
\end{abstract}

\maketitle

\section{INTRODUCTION}

Secret sharing is an important technique used for secret management .Securing secret key is very important for the proper execution of security protocols.The key idea comes from the problem of secure key storage by Shamir \cite{shamir1979} and Blakley \cite{blakley1979}, however secret sharing schemes have found numerous other applications in cryptography, secret key agreement,visual cryptography, threshold encryption, distributed computing  etc \cite{desmedt1992shared} \cite{naor1998access} \cite{ben1988completeness} \cite{chaum1988multiparty} \cite{cramer2000general} \cite{goyal2006attribute} \cite{bethencourt2007ciphertext}\cite{tassa2011generalized} \cite{shankar2008alternative}\cite{naor1995visual}.

\vskip 2mm

Both Shamir's and Blakley's proposals are $(t,n)$ threshold secret sharing schemes. The secret is shared among $n$ users and $t$ or more users can recover the secret by pooling their shares.Different proposals of threshold schemes are made using linear algebra, number theory, matroids, block codes etc \cite{asmuth1983} \cite{karnin1983} \cite{massey1993minimal} \cite{mignotte1983} \cite{kothari1985generalized} \cite{brickell1989some} \cite{simmons1992}.But Shamir's scheme was the most prominent because it offers perfect security and is flexible.The scheme is based on Lagrange Interpolation. There are also efficient $O(nlog^2n)$ algorithms which can be used for the easy implementation of Shamir's scheme.

\vskip 2mm

Extended capabilities are added to threshold schemes in the later stages.A generalized secret sharing scheme is one where any authorized subset of participants is able to recover the shared secret by pooling their shares.These authorized subset is called the access structure of the scheme. The most efficient and easy to implement scheme was Ito, Saito, Nishizeki's construction \cite{ito1989secret}.The major issue with generalized secret secret sharing scheme is the share size or the number of shares each participant has to store corresponds to each authorized access structure.Some of the proposals for the generalized secret sharing can be found in \cite{benaloh1990generalized} \cite{brickell1989some} \cite{stinson2000unconditionally} \cite{jackson1993cumulative} \cite{long2006generalised}.

\vskip 2mm

Cheater identification and detection, share verification are the major security requirement in secret sharing scheme.Each participant must be able to check the validity of the shares submitted by other participants during the reconstruction phase and also the shares distributed by the Dealer in the share distribution phase.Verifiable and Publicly Verifiable Secret Sharing(PVSS) schemes are proposed in this regard where not only the participant but any one can publicly verify the validity and consistency of the shares distributed by the Dealer.Dynamic ,proactive secret sharing etc made the schemes more flexible and error free.Tompa and Wall stated that Shamir's scheme is not cheating resistant and they proposed a scheme for detecting cheaters \cite{tompa1989share}.Signature based schemes are also proposed for identifying cheaters \cite{rabin1983randomized}.Verifiable Secret Sharing(VSS) is proposed by Benny Chor et al \cite{ben1988completeness}.Several non interactive and interactive schemes are proposed for verification \cite{ben1988completeness} \cite{feldman1987practical} \cite{pedersen1992non}.Schemes for the identification of cheaters are also proposed \cite{harn2009detection} \cite{wu1995cheating}.Publicly Verifiable Secret Sharing schemes are proposed by Stadler \cite{stadler1996publicly}. Several improvements are suggested in  \cite{fujisaki1998practical} \cite{schoenmakers1999simple}.An information theoretic secure PVSS is proposed in \cite{tang2004non}.The use of Elliptic Curve and  Pairing for PVSS is proposed by \cite{wu2011pairing}.

\vskip 2mm

Multi secret sharing is a new idea where several secrets are shared.Multiple secrets are shared among several users and when the authorized set of participant collate, they will be able to recover multiple secrets.This provides more flexibility. Each participant will hold a single share and the same share can be used to retrieve multiple secrets.The shared secrets and the threshold can also be changed dynamically but the same share can be used for the retrieval of multi secrets.Identification of cheaters and public verification of the shares make multi secret sharing scheme suitable for several applications in cloud.There are several proposals in the literature dealing with threshold multi secret sharing and generalized multi secret sharing.Most of the schemes make use of a public notice board where the dealer will publish  public parameters needed for the secret reconstruction and verification.Multiple secrets are reconstructed using the  shares of the participants and also using the global public values from the notice board.

\vskip 2mm

Karnin, Greene and Hellman \cite{karnin1983} in 1983 mentioned the  multiple secret sharing scheme.The scheme can be used to share a large secret by splitting it into smaller shares.Franklin et al \cite{franklin1992communication}, in 1992 used a technique in which the polynomial-based single secret sharing is modified, where multiple secrets are kept hidden in a single polynomial.These schemes are not perfect and are also one time use scheme. That is once the shares are used for secret reconstruction, they are exposed and cannot be used again. He and Dawson \cite{he1995multisecret} suggested a  multistage secret sharing scheme in 1994.The implementation is based on Shamir's threshold scheme and  a one way function which is hard to invert.The secrets are reconstructed stage by stage in particular order.This scheme needs $k \times n$ public values corresponds to the $k$ secrets.In 1995 Harn \cite{harn1995efficient} came up with an  alternative implementation of multi stage secret sharing which requires only $k \times (n-t)$ public values.In 2000, Chien et al \cite{chien2000practical} proposed a $(t, n)$ multi-secret sharing scheme based on the systematic block codes. Yang et al \cite{yang2004t} proposed an alternative scheme based on Shamir's secret sharing in 2004 ( YCH scheme ), which reduce the complexity of the secret reconstruction in Chien's scheme. But there are more public values required in Yang's scheme than in Chien's scheme when $p < t$. Motivated by these concerns, a new $(t, n)$ multi-secret sharing scheme is proposed by Pang and Wang \cite{pang2005new} ,in 2004 .The scheme is as easy as Yang's scheme in the secret reconstruction and requires same number of public values as Chien's scheme.Chao-Wen Chan et al \cite{chan2005scheme} proposed a multi-secret sharing scheme based on CRT and polynomial.Verifiability in multisecret sharing is implemented in\cite{shao2005new},\cite{zhao2007practical},\cite{dehkordi2008efficient}.Hash function based multi-secret sharing are proposed recently by Javier Herranz et al \cite{herranz2013new} and Jun Shao \cite{shao2014efficient}.

\vskip 2mm

Elliptic curves were found numerous applications in cryptography \cite{miller1986use}. Developed as a public key crypto system, it is found more secure with small key size compared with other public key crypto system.The Elliptic Curve Discrete Logarithm Problem (ECDLP) is much harder compared with the Discrete Logarithm Problem(DLP).The Elliptic curve group provides better security with smaller key size so the computational cost can be reduced while maintaining the same level of security.In 1993 Meneze's et al \cite{menezes1993reducing} introduced pairing.Pairing is introduced to show an attack on elliptic curve discrete logarithm problem and later found useful applications.Pairing on elliptic curve have found useful applications in identity based encryption, threshold cryptography and signature schemes, multi party key exchange etc \cite{dutta2004pairing}.The use of elliptic curve and pairing have found applications in secret sharing schemes very recently.Several schemes based on threshold and generalized secret sharing is proposed and they have found useful applications.Pairing can be used to  introduce verifiability in secret sharing scheme with more security.

\vskip 2mm

Chen Wei et al \cite{wei2007new}  proposed a dynamic threshold  secret sharing scheme using bilinear maps in 2007.Each participant holds a permanent private key.The specified threshold is realized by  adjusting the number of linear equations.The scheme is also having cheating detection capability.It is a single secret sharing scheme.A threshold multi secret sharing scheme based on ECDLP is proposed by Runhua Shi et al.\cite{shi2007t} in 2007.A fast multi-scalar multiplication scheme is also introduced.Sharing multiple secrets which are represented as points on elliptic curve using self pairing is proposed by  Liu et al \cite{liu2008new} in 2008.The proposed scheme is based on Liu et al scheme.Chen's scheme is modified to share multiple secrets by S.J.Wang et al \cite{wang2011verifiable} in 2009.In Wang's et al scheme, the number of secrets must be less than or equal to the threshold and also more public values must be modified when the secret need to be updated.Eslami et al \cite{eslami2012new} in 2010 proposed a modified scheme which avoids these problems.

\vskip 2mm

Several publicly verifiable secret sharing schemes are proposed based on pairing, but most of them are single secret sharing schemes. Youliang Tian et al \cite{tian2008practical} in 2008 proposed a  practical publicly verifiable secret sharing scheme.A pairing based PVSS is introduced by Wu and Tseng \cite{wu2011pairing} in 2011.An efficient verifiable secret sharing scheme is proposed by Jie Zhang et al \cite{zhang2014efficient}.Tian et al \cite{tian2012publicly} mentioned a distributed PVSS scheme.An efficient One Stage Multi Secret Sharing(OSMSS) is proposed recently in 2014 by Fatemi et al, \cite{fatemi2014efficient}.The scheme makes use of bilinear pairing. The number of public values are also reduced and the scheme is comparatively more efficient.

\vskip 2mm

In section 2 details about elliptic curve pairing is mentioned.Section 3 explains the Liu et al  \cite{liu2008new} point sharing algorithm on elliptic curve.Section 4 explains the modified algorithm. Experimental results and conclusions are given in section 5 and 6.

\section{ELLIPTIC CURVE AND PAIRING}

Elliptic curves appear in the implementation of many diverse areas of  cryptographic protocols.The discrete logarithm problem is more harder in elliptic curve field.The Elliptic Curve Cryptography(ECC) has found applications in devices with low power and memory.The Elliptic curve cryptography was introduced by Koblitz \cite{koblitz1987elliptic} and Miller \cite{miller1986use}.An efficient implementation ECC is proposed by Menezes et al \cite{menezes1993elliptic}.

An elliptic curve $E$ defined over $GF(q)$ is  given by $y^2=x^3+Ax+B$, where $\Delta=4A^3+27B^2$ is non zero.This ensures that the curve is non singular and has distinct roots.By adding a additional point \textit{point at infinity} $(\mathcal{O})$ it forms an additive algebraic group.So $E$ is given by $$E=\{(x,y):y^2=x^3+Ax+B\} \cup \{\mathcal{O}\}$$
The addition law makes points on elliptic curve into a commutative group.

Suppose we want to add two points $P_1=(x1,y1)$ and $P_2=(x2,y2)$.Let the line connecting $P_1$ and $P_2$ to be $$L:y=\lambda x + v$$The slope and $y$ intercept of the line is given by 
$$
\lambda= 
\begin{cases}
\frac{y_2-y_1}{x_2-x_1},& \text{if }\; P_1 \neq P_2.\\
\frac{3x_1^2+A}{2y_1}   ,& \text{if} \;P_1 = P_2.\\
\end{cases}
$$
$$v=y_1-\lambda x_1.$$
and let $$x_3=\lambda^2-x_1-x_2. \; \text{and} \; y_3=\lambda(x_1-x_3)-y_1.$$ 
Then $P_1+P_2=(x_3,y_3)$.
Elliptic curves are applied to cryptography by considering the points whose coordinate belongs to a finite field $\mathbb{F}_p$.We can define an elliptic curve $E:Y^2=X^3+AX+B$ with $A, B \in \mathbb{F}_p$ and then look for the points on $E$ with coordinates in $F_p$. The addition law satisfies all the properties and makes $E(\mathbb{F}_p)$ into a field.The Discrete Logarithm Problem(DLP) in $\mathbb{F}_p$, for a cyclic multiplicative group can be applied to Elliptic curve additive group.The Elliptic Curve Discrete Logarithm Problem(ECDLP) is defined as follows.
\begin{defn}
	Let $E$ be an elliptic curve over the finite field.Define two points  $\mathbb{F}_p$ and  $P$ and $Q$  in $E(\mathbb{F}_p)$. The ECDLP is the problem of finding an integer $n$ such that $Q=nP$.The integer $n$ is denoted by $$n=log_P(Q)$$ and $n$ is the elliptic curve discrete logarithm of $Q$ with respect to $P$.
\end{defn}

The index calculus method having sub exponential running time solves the DLP in $\mathbb{F}_p$.However there are no known general algorithms that solve ECDLP in fewer than $O(\sqrt{p})$ steps.That is ECDLP is much more difficult than the DLP.

\vskip 2mm

Pairing on elliptic curves have a number of important cryptographic applications.It is first used for cryptanalysis.The discrete logarithm problem is solved by using  MOV algorithm \cite{koblitz2000state}.The pairings technique have 
since been used to construct many cryptographic systems for which no other efficient implementation is known, such as identity based encryption or attribute based encryption \cite{dutta2004pairing}.The mapping  allows development of new cryptographic schemes based on the reduction of one problem in one group to a different, usually easier problem in the other group.The first group is usually called GAP Group where the  Decisional Diffie Hellman problem (DDHP) 
\cite{boneh1998decision} is easy,but the 
Computational Diffie Hellman (CDHP) problem remains hard.
Let $G$ be a cyclic additive group generated by $P$ whose order is prime $q$.For all $a,b,c \in \mathbb{Z}_q^*$.The CDHP is, given $P, aP, bP$ compute $abP$.DDHP is defined as, given $P, aP, bP, cP$, decide whether $c=ab$ in $\mathbb{Z}_q^*$.

\vskip 2mm

The important pairing based construct is the bilinear pairing.Lets consider two groups of the same prime order $q$ and the DLP is hard in both the groups.A mapping from $(G_1 = <P>, +)$ to $(G2,.)$ is called Bilinear Maps, if the following conditions are satisfied.
\begin{enumerate}
	\item \textbf{Bilinearity}:$\forall P,Q \in G_1,\forall a,b \in Z_q^*$ $$e(aP, bQ) = e(P,Q)^{ab}$$
	\item \textbf{ Non-Degeneracy}: If everything maps to the identity, that's obviously not interesting
	$$\forall P \in G_1,P \ne 0 \Rightarrow <e(P, P)>= G_2$$
	$$(e(P,P) \text{generates} G_2)$$
	In other words:
	$$ P \ne 0 \Rightarrow e(P,P) \ne 1$$
	\item \textbf{ Computability}: $e$ is efficiently computable.i.e., there is a polynomial time algorithm to compute $e(P,Q) \in G_2$, for all $P,Q \in G_1$.
\end{enumerate}
We can find $G1$ and $G2$ where these properties hold. The Weil and Tate provides pairings  constructions in these groups.These pairing have found numerous cryptographic applications \cite{joux2002weil}.Typically, $G_1$ is an elliptic curve group and $G_2$ is a finite field. 

\vskip 2mm

A self pairing and its applications are proposed by Lee \cite{lee2004self} in 2004.The pairing which map $e:G \times G \implies G$ is called self pairing.

\vskip 2mm

Let $K$ be a field with characteristic zero or a prime $p$ and $E = E(\bar{K})$ be an elliptic curve over $\bar{K}$, where $\bar{K}$ is an algebraic closure of $K$.Consider the set of all torsion points of order $l$ that is set of points $P$ with $lP=O$.These points forms a subgroup $E_K[l]$ of $E(K)$, where $l \ne 0$.$E[l]$ can be  represented as a direct sum of two cyclic groups. $E[l] \cong Z_l \oplus  Z_l$. That is, any point in $E[l]$ can be represented as a linear combination of two generating pair $G$ and $H$.Consider the points $P = r_1G + s_1H$ and $Q =r_2G +s_2H$ in $E[l]$, where $r_1, r_2, s_1, \; \text{and} \; s_2$ are integers in $[0, l - 1]$.We can define pairing for some fixed integers, $\alpha,\beta \in [0, l- 1]$, in the following way
$$e_{\alpha,\beta} : E[l] \times E[l] \implies E[l]$$
$$e_{\alpha,\beta}:(r_1s_2-r_2s_1)(\alpha G+\beta H)$$
The trivial case when  $\alpha,\beta$ are zero  is excluded.\\

\begin{prop}
	:The pairing $e_{\alpha,\beta}:  E[l] \times E[l] \rightarrow E[l]$
	will satisfy the properties mentioned below
	\begin{enumerate}[(i)]
		\item Identity: For all $A \in $ $E[l]$,$e_{\alpha,\beta}(A,A)=\mathcal{O}$
		\item Bilinearity: For all $A,B,C \in E[l]$
		\begin{eqnarray*}
			e_{\alpha,\beta}(A+B,C) & = & e_{\alpha,\beta}(A,C)+e_{\alpha,\beta}(B,C) \\
			e_{\alpha,\beta}(A,B+C) & = & e_{\alpha,\beta}(A,B)+e_{\alpha,\beta}(A,C)
		\end{eqnarray*}
		\item Anti-symmetry: For all $A,B \in E[l],  e_{\alpha,\beta}(A,B)= -e_{\alpha,\beta}(B,A)$
		\item Non-degeneracy: If $A \in E[l],e_{\alpha,\beta}(A,\mathcal{O})=\mathcal{O}$.More over if $e_{\alpha,\beta}(A,B)= \mathcal{O}$ for all $B \in E[l]$ then $P=\mathcal{O}$.
	\end{enumerate}
\end{prop}

\section{LIU'S SCHEME}
 Liu et al\cite{liu2008new} made the first proposal on sharing points in elliptic curve.
In this scheme multiple secrets $K_1,K_2,\ldots,K_m$ are shared and are represented as points on the Elliptic curve.The scheme consist of four main steps.
\begin{enumerate}
	\item Initialization.
	\item Share Distribution.
	\item Secret Sharing.
	\item Secret Reconstruction.
\end{enumerate}
The initialization and share distribution phase need to be done only once for
a particular $(t,n)$ secret sharing scheme, where $t$ is the threshold.Secret can be dynamically changed or more secret can be added with out modifying the participants secret share.This is achieved with a public notice board where every user have the access, but only the dealer can modify the data. 

\subsection{Initialization}

It is assumed that the dealer is a trusted authority and the participants $U_1,U_2,\ldots,U_n$ are honest.In the initialization phase some public parameters are posted on the public bulletin called notice board which can be accessed by every participant.
\begin{enumerate}
	\item The Dealer(\textbf{D}) chooses an Elliptic curve $E$ over $GF(q),q=p^r$, where $p$ is a large prime such that 
	DLP and ECDLP in GF(q) is hard. \textbf{D} then choose $E[l] \subseteq E(GF(q^k))$, a torsion subgroup of large prime 
	order $l$.
	
	\item \textbf{D} chooses a generating pair $\{G,H\} \in E[l]$ , $\alpha,\beta$, which is used for pairing and compute $W=\alpha.G+\beta.H$
	
	\item  \textbf{D} then publish $\{E,q,l,k,W\}$ in the notice board.
\end{enumerate}

\subsection{Share Distribution}

In this phase, shares are generated and are distributed to $n$ participants.Any $t$ of them can reconstruct the secret. These shares are independent of the secret to be distributed.
\begin{enumerate}
	\item \textbf{D} generate a matrix $A$ of size $n \times t$ as:\\$ A=\left(\begin{array}{ccccc}
	1 & 1 & 1 & \; \;\ldots & 1\\
	1 & 2 & 2^2 & \; \;\ldots & 2^{t-1}\\
	1 & 3 & 3^2 & \;\;\ldots & 3^{t-1}\\
	\vdots  & \vdots  &  \vdots   &\; \;\vdots & \vdots\\
	1 & n & n^2& \; \;\ldots & n^{t-1}
	\end{array}\right)$
	
	\item \textbf{D} randomly choose $t$ pairs of numbers ${a_i\prime,b_i\prime} \in [1,l-1],1 \le i \le t$ and computes \\
	\begin{eqnarray*}
		(a_1,a_2,\ldots,a_n)^T  &=& A.(a_1\prime,a_2\prime,\ldots,a_t\prime)^T \\ [.1cm] 
		(b_1,b_2,\ldots,b_n)^T &=& A.(b_1\prime,b_2\prime,\ldots,b_t\prime)^T
	\end{eqnarray*}
	
	\item \textbf{D} sends $P_j=\{a_j,b_j\}$ as a secret share to each user $U_j, 1\le j \le n$ through a secure channel.
\end{enumerate}
\subsection{Secret Sharing}
After distributing the secret shares, the dealer will share multi-secret by publishing informations on a public notice board corresponding to each secret to be shared.Each participant can make use of these public values. These public information together with the secret share of each participant can be used for the retrieval of shared secrets.The number of secrets $m$ must be less than or equal to the threshold $t$ for the scheme to work.
\begin{enumerate}
	\item The secrets to be shared ie; $K_1,K_2,\ldots,K_m$ is mapped into $m$ points on the Elliptic curve $M_1,M_2,\ldots,M_m$.
	\item \textbf{D} chooses $\{c_i,d_i\} \in [0,l-1]$ randomly and computes $Q_i=c_i.G+d_i.H$ and $R_i=e_{{\alpha},{\beta}}(Q_i,P_i\prime)+M_i$, for all $1 \le i \le m$.
	\item \textbf{D} then publish $\{c_i,d_i,R_i\}$,$1 \le i \le m$, on the public bulletin.
\end{enumerate}
\subsection{Secret Reconstruction}
Let $t$ users $U_1,U_2,\ldots,U_t$ wants to reconstruct $m$ secrets.Each user contribute his pseudo share for the reconstruction of secret.The pseudo share is computed from his secret share and the public informations.

For each secret $K_i$ from 1 to $m$ and for each user $U_j$ from 1 to $t$
\begin{enumerate}
	\item Each $U_j$ download the pair $\{c_i,d_i\}$ from the public bulletin and compute pseudo share $SS_{ij}=e_{\alpha,\beta}(Q_i,P_j)$, where $P_j=a_j.G+b_j.H$ and $Q_i=c_i.G+d_i.H$, $1 \le i \le t$ and $1 \le j \le t$.
	\item Each user $U_j$ multi-casts the pseudo share $S_{ij}$ to other $t-1$ participants.
	\item Each user then computes $T_i= \sum_{k=1}^{t} y_k.S_{ij}$, where
	$y_k = \prod_{j=1,j\neq k}^{t}(k-j))^{-1}$.
	\item Each user can download the point $R_i$ from the public bulletin and recovers $M_i=R_i-T_i$.
\end{enumerate}

\section{PROPOSED SCHEME}
The major difficulties with Liu's scheme is that the secrets are represented as points on the elliptic curve.The mapping of secret to the
elliptic curve point is very difficult.The number of secrets that can be shared also depends on the threshold $t$.We cannot share more than $t$ secrets.The Dealer is assumed to be a trusted authority.There is no provision for the verification of the shares distributed by the Dealer and also the participants cannot verify the shares distributed by the other participant during the reconstruction.The proposed scheme overcome these difficulties.We make use of self pairing and bilinear pairing for the efficient construction of the scheme.

The proposed scheme is a threshold multi secret sharing scheme where any number of secrets can be shared and threshold number of users can reconstruct the multi secrets.The multi secrets can be reconstructed in single stage, unlike the multi stage secret sharing scheme where the secrets are reconstructed stage by stage.Each participant can verify their secret share, during the secret distribution phase by the dealer and also verify the shares send by other participants during the reconstruction phase to identify the cheaters.

The proposed scheme consist of the following three phases
\begin{enumerate}
	\item Initialization and Secret Sharing.
	\item Secret Reconstruction
	\item Verification
\end{enumerate}

\subsection{Initialization and Secret sharing}
In the initialization phase some public parameters are posted on the public bulletin called notice board which can be accessed by every participant.
Let $U_1,U_2 \ldots U_n$ be the $n$ users involved in the secret sharing
and let $K_1,K_2,\ldots,K_m$ be the m secrets to be shared.
\begin{enumerate}
	\item The Dealer(\textbf{D}) chooses an Elliptic curve $E$ over $GF(q),q=p^r$, where $p$ is a large prime such that DLP and ECDLP in GF(q) is hard. \textbf{D} then choose $E[l] \subseteq E(GF(q^k))$, a torsion subgroup of large prime order $l$.
	
	\item \textbf{D} chooses a generating pair $\{G,H\} \in E[l]$  and $\alpha,\beta \in [1,l-1]$, which is used for pairing and compute $W=\alpha.G+\beta.H$
	
	\item  \textbf{D} then publish $\{E,q,l,k,G,H,W\}$ in the notice board.
	
	\item A secret point $P_0 $ is chosen where $P_0=a_0.G+ b_0.H$
	\item The Dealer then construct a polynomial of degree $t-1$ .
	$$f(x)=a_0+a_1x^1+a_2x^2+\cdots+a_{t-2}x^{t-2}+ b_0x^{t-1}$$ where $a_0$ and $b_0$ corresponds to the point $P_0$ and the other coefficient values are chosen from ${\mathbb{Z}_l}^*$
	\item $D$ compute shares $S_i=f(x_i) (\mod l)$ ,and send the shares $P_i=(x_i,S_i)$	secretly to the  users $U_i$ for $i=1,\ldots,n.$
	
	\item The dealer also publishes the values $c,d$ corresponds to a point
	$Q=c.G+d.H$	and the verification point $V_i=e_{\alpha,\beta}(P_i,Q)$ for $i=1,\ldots,n$ corresponds to each share and also $V_0=e_{\alpha,\beta}(P_0,Q)$.
	
	\item Publish the recovery code $R_i=K_i-e(P_0,i.P_0)$ for $i=1,\ldots ,m$, corresponds to each secret that is to be shared.
\end{enumerate}
\subsection{Secret Reconstruction}
\begin{enumerate}
	\item When the threshold number of users want to reconstruct the secret , they pool the shares and reconstruct the polynomial $f(x)$ using Lagrange Interpolation.
	$$f(x)=\sum_{j=1}^{t} S_j \prod_{1 \le i \le t,i \ne j}\frac{x-x_i}{x_j-x_i}$$
	
	\item From the reconstructed polynomial $a_0$ and $b_0$ can be obtained and hence $P_0$ can be obtained.
	
	\item Using the published recovery codes , the $m$ secrets can be recovered by
	$$K_i=R_i+e(P_0,i.P_0)$$ for $i=1,\ldots,m$
\end{enumerate}

\subsection{Verification}
Each user can verify the shares using the share verification point $V_i$.
\begin{enumerate}
	\item Each user compute $v_i=e_{\alpha,\beta}(P_i,Q)$ using his share and the public value $Q$
	\item If the computed value $v_i=V_i$, the share send by the Dealer is valid.
\end{enumerate}

During the reconstruction stage, the validity of the shares send by each user can be verified by using the same technique mentioned above.Also the consistency of the shares are verified by checking $V_0$.That is, after all the shares are pooled , the polynomial is reconstructed.Using $a_0,b_0$,and the generators $G,H$, $P_0$ can be obtained.The participant can verify the consistency of the shares by checking $V_0=v_0$ , where $v_0=e_{\alpha,\beta}(P_0,Q)$, and $V_0$ is the corresponding published value.

\section{SECURITY ANALYSIS}
One of the major requirement of the secret sharing scheme is the consistency of the shares.
\begin{thrm}
	The probability that the Dealer distribute inconsistent shares to the participant is negligible.
\end{thrm}
\textit{\textbf{proof.}} The coefficients of the polynomial $f(x)$ are chosen by the dealer.The values of $x_i$ are chosen such that the inverse must exist.How ever the Dealer cannot send invalid shares to the participants. Because the share $(x_i,S_i)$ can be verified by each participant using self pairing with the public value $Q$.Also the consistency of the shares can be verified by checking $e_{\alpha,\beta}(P_0,Q)=V_0$.If the  shares are inconsistent, then the computed value will not match with $V_0$.

\begin{thrm}
	The probability that the participant distribute invalid shares during the reconstruction is negligible.
\end{thrm}
\textit{\textbf{proof.}} Each participant can verify the shares send by the other participants by checking $e_{\alpha,\beta}(P_i,Q)=V_i$, if there is a mismatch the participant is a cheater and we are able to detect and identify the cheater.
\begin{thrm}
	Adversary cannot derive any information about the secret from the public values.
\end{thrm}
\textit{\textbf{proof.}} The polynomial $f(x)$ is of degree $t-1$, so less than $t$ participant cannot derive any useful information about the secret by pooling their shares.The verification code $Vi=e_{\alpha,\beta}(P_i,Q)$ cannot reveal any info about the share $P_i$.The point $V_i$ is a linear combination of the generator $(G,H)$, and the coefficients can be any value from the field $\mathbb{Z}_l^*$. the public values $R_i$ also cannot give any information about the shared secret.The 
non degeneracy of pairing ensures that $e(P_0,P_0)$ is a primitive $l^{th}$ root of unity for all non zero $P_0 \in E(F_q)[l]$.
\begin{thrm}
	Finding $P_0$ is as difficult as guessing the secret.
\end{thrm}
\textit{\textbf{proof.}} The security of the system depends on finding $P_0$ and is again depends on $a_0$ and $b_0$. These values can be retrieved only by reconstructing the  $t-1$ degree polynomial by the $t$ users involved in secret reconstruction.The shares have the same size as these parameters and are elements of the same field.This provides information theoretical security. Less than $t$ participant cannot derive any useful information because of the security of  Shamir's scheme. Hence the adversary can only guess the values. The probability is $1/l$, and is negligible when $l$ is large.Finding $P_0$ without knowing $a_0$ and $b_0$ is again like trying all linear combinations of the generators $G$ and $H$, which is again a more complex process.Without knowing $P_0$ finding the secret from the public parameter $R_i$ is as complex as guessing the secret.

\section{EXPERIMENTAL RESULTS}

SAGE and Python are used for implementing the scheme.A simple example showing sharing of two secrets $K_1,K_2$ according to $(2,3)$ threshold scheme  is mentioned here.We considered field with smaller prime power for easy  understanding.
\subsection{Initialization}
\begin{enumerate}
	\item Elliptic Curve defined by $E:y^2 = x^3 + 4.x$ over Finite Field in $i$ of size $47^6$ is chosen for secret sharing.The order of $E$ is $10779422976(2^8 . 3^4 . 7^2 . 103^2)$.Additive Abelian group isomorphic to $Z/103824 + Z/103824$ is embedded in Abelian group of points on the curve.
	
	\item The dealer $D$ chooses a torsion subgroup $E[103]\subseteq E(GF(47^6))$.
	Two randomly chosen generator pairs ${G,H}$ of  $E[103]$ are
	\begin{eqnarray*}
		G =&&(19i^5 + 38i^4 + 26i^3 + 28i^2 + 45i + 6 :\\ [0.1cm]
		&& 20i^5 + 18i^4 + 12i^3 + 32i^2 + 12i + 43 : 1) \\ [0.1cm]
		H =&&(5i^5 + 8i^4 + 41i^3 + 46i^2 + 39i + 34 :  \\ [0.1cm]
		&& 32i^5 + 7i^4 + 18i^3 + 34i^2 + 8i + 32 : 1)
	\end{eqnarray*}
	
	Let $\alpha=51,\beta=35$ , compute $W=\alpha.G+\beta.H$ 
	\begin{eqnarray*}
		W=&&(25i^5 + 3i^4 + 11i^3 + 15i^2 + 39i + 19 : \\ [0.1cm]
		&& 40i^5 + 41i^4 + 9i^3+ 44i^2 + 22i + 1 : 1)
	\end{eqnarray*}
	
	\item  $E:y^2 = x^3 + 4.x,q=47^6,l=103,k=1,\alpha=51,\beta=35,W=(25i^5 + 3i^4 + 11i^3 + 15i^2 + 39i + 19 : 40i^5 + 41i^4 + 9i^3+ 44i^2 + 22i + 1 : 1)$ are made public.
\end{enumerate}
\subsection{Share Distribution}
\begin{enumerate}
	\item The matrix $A$ for a $(2,3)$ scheme is of size $3 \times 2$
	\begin{equation*}
		A=\left(\begin{array}{rr}
			1 & 1 \\
			1 & 2 \\
			1 & 3
		\end{array}\right)
	\end{equation*}
	\item Dealer chooses two pairs of random numbers from $[1,102]$ 
	$a_1\prime=11,a_2\prime=25,b_1\prime=15,b_2\prime=33$ and compute
	\begin{eqnarray*}
		A \times [a_1\prime,a_2\prime]&=[36,61,86]&=[a_1,a_2,a_3]\\[0.1cm]
		A \times [b_1\prime,b_2\prime]&=[48,81,114]&=[b_1,b_2,b_3]
	\end{eqnarray*}
	
	\item The three users  $U_1,U_2,U_3$ will get shares $P_1=(36,48),P_2=(61,81),P_3=(86,114)$.
\end{enumerate}
\subsection{Secret Sharing}
\begin{enumerate}
	\item We consider two secrets $K_1$ and $K_2$ to be shared.These secrets are mapped into Elliptic curve points $M_1$ and $M_2$.
	\begin{eqnarray*}
		M_1=&(19i^5+38i^4+26i^3+28i^2+45i+6:\\ 
		[0.1cm]
		& 20i^5+18i^4+12i^3+ 32i^2 + 12i + 43 : 1)\\ [0.1cm]
		M_2=&(5i^5+8i^4+41i^3+46i^2+39i+34:\\ 
	 [0.1cm]
		& 32i^5+7i^4+18i^3 + 34i^2+8i+32:1)
	\end{eqnarray*}
	
	\item In order to share the two secrets,\textbf{D} chooses two random pairs of numbers $(c_1,d_1),(c_2,d_2) \in [0,l-1]$ 
	and compute $Q_i=c_i.G+ d_i.H$ for the two secrets.Dealer computes pairing $R_i=e_{\alpha,\beta}(Q_i,P_i \prime )+M_i$ corresponds to each secret, where $P_i\prime=a_i\prime.G+b_i\prime.H$. 
	
	Let $c_1=15,d_1=11,c_2=23,d_2=39 \; \mbox{and} \; a_1\prime=11,b_1\prime=15,a_2\prime=25,b_2\prime=33$.
	\begin{eqnarray*}
		R_1 &=& e_{51,35}(Q_1,P_1\prime)\\ [0.1cm]
		R_1&=&(c_1b_1\prime-d_1a_1\prime)(\alpha G+\beta H)+M_1\\ [0.1cm]
		R_1&=&(15.15-11.11).W+M_1\\ [0.1cm]
		R_1&=&(i^5 + 27i^4 + 5i^3 + 7i^2 + 35i + 38 : \\ [0.1cm]
		&& 21i^5 + 44i^4 + 28i^3 +15i^2 + 9i + 16 : 1)\\ [0.1cm]
		\ \\ [0.3cm]
		R_2 &=& e_{51,35}(Q_2,P_2\prime)\\ [0.1cm]
		R_2&=&(c_2b_2\prime-d_2a_2\prime)(\alpha G+\beta H)+M_2\\ [0.1cm]
		R_2&=&(23.33-25.39).W+M_2 \\ [0.1cm]
		R_2&=&(33i^5 + 43i^4 + 20i^3 + 17i^2 + 39i + 33 : \\ [0.1cm]
		&& 45i^5 + 5i^4 + 43i^3+ 24i^2 + 41i + 38 : 1)
	\end{eqnarray*}
	
	\item Publish 
	\begin{eqnarray*}
		\{c_1,d_1,R_1\} &=& \{15,11,(i^5 + 27i^4 + 5i^3 + 7i^2 \\ [0.1cm]
		&&+ 35i + 38 : \\ [0.1cm]
		&& 21i^5 + 44i^4 + 28i^3 +15i^2 \\ [0.1cm]
		&&+ 9i + 16 : 1)\}\\ [0.1cm]
		\{c_2,d_2,R_2\} &=& \{23,39,(33i^5 + 43i^4 + 20i^3 + 17i^2\\ [0.1cm]
		&&+ 39i + 33 : \\ [0.1cm]
		&& 45i^5 + 5i^4 + 43i^3+ 24i^2 + \\ [0.1cm]
		&& 41i + 38 : 1)\}
	\end{eqnarray*}
\end{enumerate}
\subsection{Secret Reconstruction}
Assume that participants $P_1$ and $P_2$ want to reconstruct the secrets
$K_1,K_2$. 
\begin{enumerate}
	\item Each participant compute his share contribution for the reconstruction of each secret as $$S_{ij}=e_{\alpha,\beta}(Q_i,P_i)$$
	\begin{eqnarray*}
		S_{11}&=&  e_{51,35}(Q_1,P_1)\\[0.1cm]
		S_{11}&=& (c_1.b_1-d_1.a_1).W\\[0.1cm]
		S_{11}&=& (15.48-11.36).W\\[0.1cm]
		S_{11}&=& (8i^5 + 5i^4 + 2i^3 + 29i^2 + 13i + 7 :\\ [0.1cm]
		&&  4i^5 + 29i^4 + 44i^3 + 43i^2 + 6i + 20 : 1)  
		\ \\ [0.3cm]
		S_{12}&=& e_{51,35}(Q_1,P_2)\\ [0.1cm]
		S_{12}&=& (15.81-11.61).W\\[0.1cm]
		S_{12}&=&(37i^5 + 28i^4 + 31i^3 + 15i^2 + 40i + 44 :\\[0.1cm]
		&& 38i^5 + 5i^4 + 3i^3 + 39i^2 + 26i + 8 : 1) \\[0.1cm]
		S_{21}&=&(5i^5 + 36i^4 + 31i^3 + 34i^2 + 5i + 38 :\\[0.1cm]
		&& 5i^5 + 22i^4 + 13i^3 +4i^2 + 39i + 17 : 1)\\[0.1cm]
		S_{22}&=&(25i^5 + 3i^4 + 11.i^3 + 15i^2 + 39i + 19 :\\[0.1cm]
		&& 7i^5 + 6i^4 + 38i^3 + 3i^2 + 25i + 46 : 1)
	\end{eqnarray*}
	\item The shares generated are then multi-casted.
	\item The inverse of the matrix A is 	
	$\left(\begin{array}{rr}
	2 & -1 \\
	-1 & 1
	\end{array}\right)$
	.Each user then compute $y_k.S_{ij}$.
	\begin{eqnarray*}
		T_1&=& 2S_{11}-1S_{12} \\[0.1cm]
		T_1&=& (25i^5 + 3i^4 + 11i^3 + 15i^2 + 39i + 19 :\\[0.1cm]
		&& 40i^5 + 41i^4 + 9i^3+ 44i^2 + 22i + 1 : 1)\\[0.1cm]
		T_2&=& -1S_{21}+1S_{22}\\[0.1cm]
		T_2&=& (29i^5 + 43i^4 + 19i^3 + 20i^2 + 36i + 25 :\\ [0.1cm]
		&& 45i^5 + 9i^4 + 29i^3+ 15i^2 + 9i + 31 : 1)\\[0.1cm]
	\end{eqnarray*}
	\item Each participant can download $R_i$ from the public bulletin and reconstruct $M_i=R_i-T_i$.
	\begin{eqnarray*}
		M_1 &=& R_1-T_1 \\[0.1cm]
		M_1 & = & (19i^5 + 38i^4 + 26i^3 + 28i^2 + 45i + 6 : \\ [0.1cm]
		& & 20i^5 + 18i^4 + 12i^3+ 32i^2 + 12i + 43 : 1) 
		\ \\[0.3cm]
		M_2 & = & R_2-T_2 \\[0.1cm]
		M_2 & = &(5i^5 + 8i^4 + 41i^3 + 46i^2 + 39i + 34 :\\[0.1cm]
		& & 32i^5 + 7i^4 + 18i^3 +34i^2 + 8i + 32 : 1) 
	\end{eqnarray*}
\end{enumerate}
From $M_1$,$M_2$,$K_1$ and $K_2$ can be obtained.

\vskip 2mm

The comparison of various schemes which uses elliptic curve and bilinear pairing is studied and is shown in Table \ref{tab:comp}.We considered recent proposal of only the threshold multi secret sharing schemes based on elliptic curve and pairing.Lets consider a $(t,n)$ threshold multi secret sharing scheme which can share $m$ secrets.It is found that in Liu's\cite{liu2008new} and Wang's\cite{wang2011verifiable} scheme, the number of secret that can be shared is proportional to $t$.So these schemes are not suitable for sharing more than $t$ secrets.Wangs scheme is a modification of Chen's single secret sharing scheme.Eslami again modified the Wang's scheme .The advantage of our proposed scheme is that, large number of secret can be shared and also the public values used are less.Most of the schemes mentioned in the literature having the share verification property.But they use discrete logarithm problem.This verification code can reveal information about the secret if discrete logarithm problem is tractable. Hence the security of the shared secret also depends on the hardness of the DLP problem.This is avoided in our scheme.The verification code does not reveal any information about the secret and is more secure compared with the existing scheme.
\begin{table*}[t]
	\renewcommand{\baselinestretch}{1}
	\caption {comparison of various schemes using elliptic curve and pairing}
	\label{tab:comp}
	\begin{small}
		\begin{center}
	\begin{tabular}{|l|l|l|l|l|l|l|}
		\hline scheme 				& Liu \cite{liu2008new}	 & Chen  \cite{wei2007new} & Wang  \cite{wang2011verifiable} & Eslami  \cite{eslami2012new} & Proposed \\ 
		\hline  single(ss)/multi secret(ms)  & ms & ss &  ms& ms & ms \\ 
		\hline secrets & t & 1 & t & m +n & $m \ge n$ \\
		\hline public parameters 	 & 5 + 3m & 8 +n-t & 8+2*n & 8+n+m-t & 7+n+m \\  
		\hline single stage 			     & Yes & Yes &   Yes & Yes& Yes \\ 
		\hline verifiability 				 &  No & Yes &   Yes& Yes & Yes \\ 
		\hline cheater detection 			 & No & Yes  &   No& Yes &  Yes\\ 
		\hline cheater identification 		 & No & Yes  &  No & Yes & Yes\\ 
		\hline
	\end{tabular} 
\end{center}
\end{small}
\end{table*}

\section{CONCLUSIONS}
In this paper we have proposed a novel threshold multi-secret sharing scheme based on elliptic curve and bilinear pairing .Most of the scheme proposed in the literature use bilinear pairing for verification of shares or identification of cheating.We have used the method of point sharing and verification using self pairing.A non degenerate Tate pairing or modified Weil pairing can be used to share multiple secrets.The number of public parameters are greatly reduced and the security does not depend on the hard  computational  problem.The verification mechanism can prevent users from cheating. Also the consistency of the shares can be verified by the participants which avoids the need of a trusted dealer.The proposed scheme is the first multi secret sharing scheme with the extended capabilities of share verification and cheater identification based on self pairing.The use of elliptic curve and self pairing can be further explored to develop secret sharing schemes with more generalized access structure.

\bibliographystyle{abbrv}
\bibliography{ss}

\noindent{\includegraphics[width=1in,height=1.7in,clip,keepaspectratio]{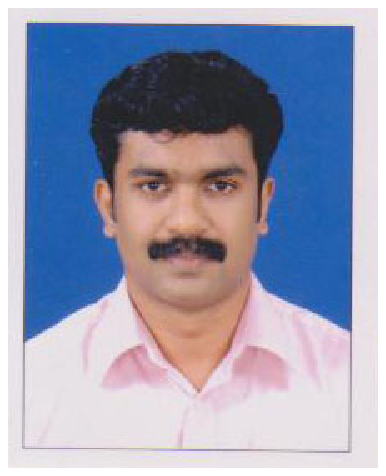}}
\begin{minipage}[b][1in][c]{1.8in}
	{\centering{\bf {Binu V P}} is currently a  Research Scholar in the Department of Computer Applications, Cochin University of Science and Technology(CUSAT). He Holds a Bachelor Degree in Computer Science and Engineering and Masters Degree  in Computer } 
\end{minipage}
\vskip 3mm
and Information Science. His research area includes Cryptography and Secret Sharing.\\\\
\noindent{\includegraphics[width=1in,height=1.7in,clip,keepaspectratio]{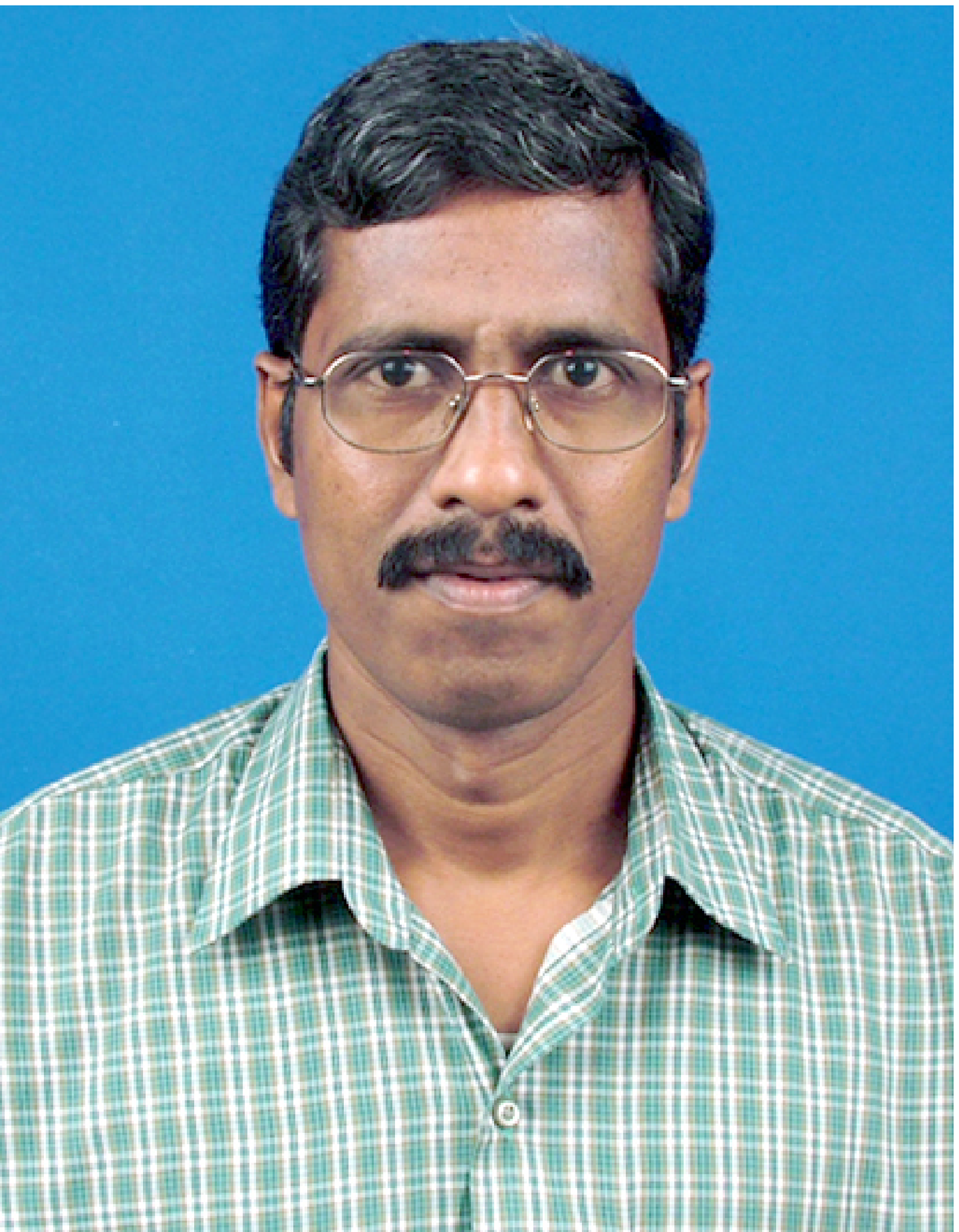}}
\begin{minipage}[b][1in][c]{1.8in} 
	{\centering{ \bf{A Sreekumar}} received his MTech Degree in Computer Science and Engineering from IIT Madras, in 1992 and Ph.D in Cryptography from Cochin University of Science and Technology, in 2010. He joined as a Lecturer in the Department  of Computer Applications, CUSAT} \\\\
\end{minipage}
\vskip 2mm
in 1994 and currently he is working as an Associate Professor. He had more than 20 years of teaching experience. His research interest includes Cryptography, Secret Sharing Schemes and Number Theory.
\small\balance
\end{document}